\documentclass[twocolumn,showpacs,superscriptaddress,prl]{revtex4-1}

\usepackage[dvips]{graphicx}
\usepackage{dcolumn}			
\usepackage{color}
\usepackage{ulem}
\usepackage{amsmath}
\begin{document}

\title{Rotation of the magnetic vortex lattice in Ru$_{7}$B$_{3}$ driven by the effects of broken time-reversal and inversion symmetry}

\author{A. S. Cameron}
\affiliation{Institut f\"ur Festk\"orper- und Materialphysik, Technische Universit\"at Dresden, D-01069 Dresden, Germany}
\author{Y. S. Yerin}
\affiliation{Institut f\"ur Festk\"orper- und Materialphysik, Technische Universit\"at Dresden, D-01069 Dresden, Germany}
\affiliation{Physics Division, School of Science and Technology, Universit\`{a} di Camerino, Via Madonna delle Carceri 9, I-62032 Camerino (MC), Italy}
\author{Y. V. Tymoshenko}
\author{P.~Y.~Portnichenko}
\affiliation{Institut f\"ur Festk\"orper- und Materialphysik, Technische Universit\"at Dresden, D-01069 Dresden, Germany}
\author{A. S. Sukhanov}
\affiliation{Max Planck Institute for Chemical Physics of Solids, D-01187 Dresden, Germany}
\affiliation{Institut f\"ur Festk\"orper- und Materialphysik, Technische Universit\"at Dresden, D-01069 Dresden, Germany}
\author{M. Ciomaga Hatnean}
\author{D. McK. Paul}
\author{G. Balakrishnan}
\affiliation{Department of Physics, University of Warwick, Coventry, CV47AL, United Kingdom}
\author{R. Cubitt}
\affiliation{Institut Laue-Langevin, 71 avenue des Martyrs, CS 20156, F-38042 Grenoble Cedex 9, France}
\author{D. S. Inosov}
\affiliation{Institut f\"ur Festk\"orper- und Materialphysik, Technische Universit\"at Dresden, D-01069 Dresden, Germany}

\begin{abstract}
We observe a hysteretic reorientation of the magnetic vortex lattice in the noncentrosymmetric superconductor Ru$_{7}$B$_{3}$, with the change in orientation driven by altering magnetic field below $T_{\rm c}$. Normally a vortex lattice chooses either a single or degenerate set of orientations with respect to a crystal lattice at any given field or temperature, a behavior well described by prevailing phenomenological and microscopic theories. Here, in the absence of any typical VL structural transition, we observe a continuous rotation of the vortex lattice which exhibits a pronounced hysteresis and is driven by a change in magnetic field. We propose that this rotation is related to the spontaneous magnetic fields present in the superconducting phase, which are evidenced by the observation of time-reversal symmetry breaking, and the physics of broken inversion symmetry. Finally, we develop a model from the Ginzburg-Landau approach which shows that the coupling of these to the vortex lattice orientation can result in the rotation we observe.
\end{abstract}

\date{\today}

\maketitle

\section{Introduction}

Interest in noncentrosymmetric (NCS) superconductors has greatly increased since the discovery of the heavy-fermion superconductor CePt$_{3}$Si~\cite{Bau04}, and many novel superconducting states with unusual properties have been predicted. The key physics in NCS superconductors is that of antisymmetric spin-orbit coupling (ASOC), which spin-splits the Fermi surface, removes the conservation of parity, and permits the mixing of \textit{s}- and \textit{p}-wave states~\cite{Sig91, Gor01, Fig04, Sig07}. Singlet-triplet mixing has only been observed in some cases, as both must be allowed by the pairing mechanism, and the ASOC must be strong enough for the effects to become noticeable. Perhaps the best known example is the case of Li$_{2}$Pd$_{3}$B$_{3}$ and Li$_{2}$Pt$_{3}$B$_{3}$, where the Pd system appeared to have a predominantly spin-singlet order parameter while the larger spin-orbit coupling in the Pt system resulted in a dominant triplet component, and thus line nodes in the energy gap as evidenced by penetration depth measurements~\cite{Yua06}.

An order parameter consisting of a singlet-triplet mixture should strongly affect the electronic states around a vortex core~\cite{Kau05, Nag06}, and can potentially introduce nodes in the gap not demanded by symmetry~\cite{Hay06, Fri06}. Vortex core anisotropies and nodal gaps are well known to result in structural phase transitions of the vortex lattice (VL) as the applied magnetic field and temperature are varied~\cite{Lav10, Suz10}, and as such the VL may be an ideal probe to investigate broken inversion symmetry. Thus, VL structure transitions are not unusual, having been observed in classical superconductors~\cite{Lav10}, cuprates~\cite{Gil02, Whi09, Whi14, Cam14}, pnictides~\cite{Furukawa11, Mor14} and others~\cite{Esk97}, to name but a few. In theory, these transitions are generally described as resulting from anisotropy in either the superconducting gap~\cite{Ich99}, Fermi velocity~\cite{Kog97a, Kog97b, Fra97} or both~\cite{Nak02,Suz10} and are driven by thermal fluctuations at a transition line in the manner of a classical phase transition. VL structures which show a gradual field and temperature dependence are also known, which can be driven by the same physics as the structural transitions described above, but also by multi-gap physics in the case of MgB$_{2}$, where the VL undergoes a smooth rotation as a function of field~\cite{Das12, Hir13}. While, to date, the theories focusing on gap and Fermi velocity anisotropy have not been adapted for NCS superconductors, the effect of broken inversion symmetry on the VL has been the subject of multiple studies \cite{Nag06, Yip05, Kas13, Mat08, Hia09}, which have focused on the $C_{4v}$ and $O$ crystallographic point groups. Perhaps the most striking result from these investigations has been the appearance of a transverse component of magnetic field in the vortex lattice \cite{Yip05, Mas06, Hay06, Lu08, Lu09, Kas13}, which arises due to currents flowing parallel to the vortex which are unique to NCS systems. Further, the emergence of a new gap-amplitude modulated phase has been predicted in superconductors with non-zero Rashba-type spin-orbit coupling, which should have a strong effect on the VL coordination~\cite{Mat08, Hia09}, although to date neither of these has been directly observed. To our knowledge, the only NCS superconductor where the VL morphology has been studied is BiPd~\cite{Lem16}, which displayed an intermediate mixed state but otherwise showed no signs of unconventional behavior related to broken inversion symmetry.

Here we employ small-angle neutron scattering (SANS) to study the VL in another NCS superconductor, Ru$_{7}$B$_{3}$. It forms a NCS crystal structure with the space group $P6_{3}mc$~\cite{Aro59}, which is hexagonal in the basal plane. Our single-crystal sample has a superconducting transition temperature of $T_{\rm c} = 2.6$~K~\cite{Sin14}, which sits within the range of 2.5 to 3.4~K observed in earlier studies~\cite{Mat61, Fan09, Kas09}. It is reported to have an isotropic \textit{s}-wave gap~\cite{Fan09}, rather than the singlet-triplet mixture predicted for NCS superconductors. Specific heat and magnetization measurements on a single crystal of Ru$_{7}$B$_{3}$ resulted in Ginzburg-Landau parameters of 21.6 and 25.5 for the $[100]$ and $[001]$ directions respectively~\cite{Kas09}, making it a reasonably strong type-II superconductor (in contrast to BiPd).

\section{Experimental details}

SANS measurements were performed on the D33 instrument~\cite{Dew08} at the Institut Laue Langevin in Grenoble, France~\cite{ILL15, ILL16}. Incoming neutrons were velocity selected with a wavelength between 10 and 14~\AA, depending on the measurement, with a FWHM in wavelength spread of $\sim 10 \%$, and diffracted neutrons were detected using a position sensitive detector. The sample was mounted on a copper holder with the \textbf{a} and \textbf{c} directions in the horizontal plane and placed in a dilution refrigerator within a horizontal-field cryomagnet with the magnetic field applied along the neutron beam. Since $T_{\rm c}$ was above the maximum stable temperature of the dilution refrigerator, the sample was cooled in no applied field (zero field cooled, or ZFC), and the magnetic field was applied and changed while at base temperature. Measurements, such as those in Fig.~\ref{Fig1}, were taken by holding the applied field and temperature constant and rocking the sample throughout all the angles that fulfill the Bragg conditions for the first-order diffraction spots of the VL. Background measurements were taken in zero field and then subtracted from the in-field measurements to leave only the VL signal. Diffraction patterns were treated with a Bayesian method for handling small-angle diffraction data, detailed in Ref.~\cite{Hol14}.

\section{Results}

\begin{figure*} 
	\includegraphics[width=1\linewidth]{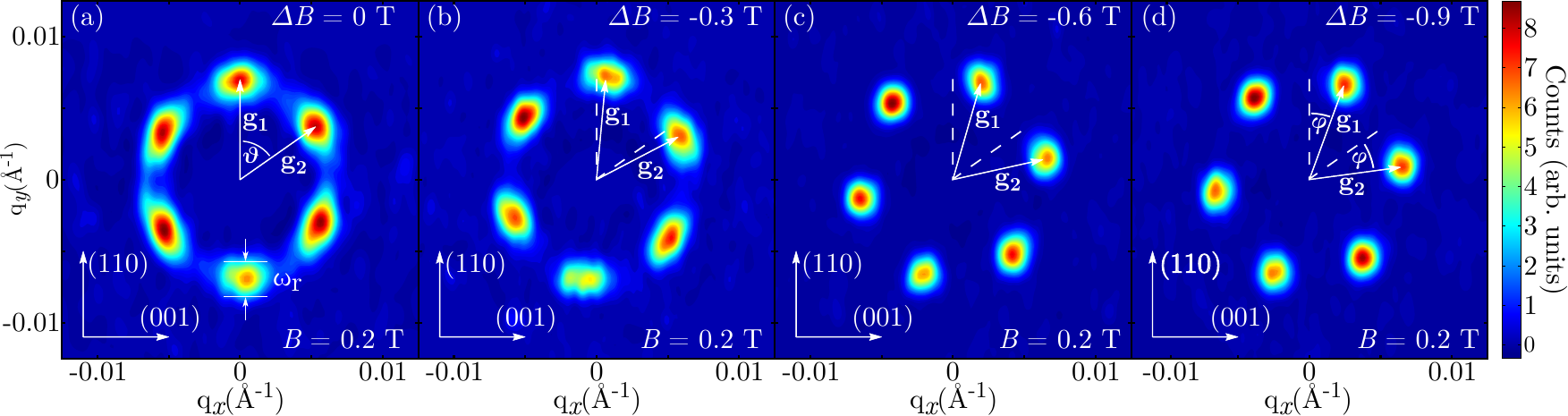}
	\caption{Diffraction patterns from the VL taken at 0.2~T along the \textbf{a}-axis after different field histories: (a) $0 \rightarrow 0.2$~T, (b) $0 \rightarrow 0.5 \rightarrow 0.2$~T (c) $0 \rightarrow 0.9 \rightarrow 0.2$~T and (d) $0 \rightarrow 1.1 \rightarrow 0.2$~T. The magnitude of the decrease in magnetic field prior to measurement is indicated as $\Delta B$. The reciprocal space lattice vectors of the VL, \textbf{g$_{1}$} and \textbf{g$_{2}$}, are shown for each diffraction pattern, and the radial spot width $\omega_{\rm r}$ is illustrated in panel (a). The angle between the two basis vectors, $\theta$, is shown in panel (a), while the orientation of the lattice with respect to the ZFC lattice, $\phi$, is shown in panel (d).}
	\label{Fig1}
\end{figure*}

Now we turn to the presentation of SANS data on Ru$_{7}$B$_{3}$. We focus on measurements at 0.2~T and above, as below this field the strength of the vortex pinning is high enough to disorder the VL, which complicates diffraction measurements. For magnetic fields applied along the \textbf{a} axis we observe a hexagonal VL with a small degree of anisotropy, on the order of $9 \%$, up to the maximum measured field of 1~T. However, we find that the orientation of the VL with respect to the crystal lattice is not simply dependent on the magnitude of the applied field, as it is common in many superconductors, but on the field history when below the critical temperature. Fig.~\ref{Fig1} shows diffraction patterns all taken at 0.2~T, and we observe the orientation of the VL change significantly depending on the field history of the sample. Panel (a) shows the VL at 0.2~T, applied from zero field at 55~mK. We consider this orientation to be the `equilibrium state'~\cite{Cam88} of the VL against which we measure other orientations, as we find the ZFC procedure repeatedly reproduces it for all of our measurable field range, and it is approximately the same orientation observed when the sample is cooled through $T_{\rm c}$ while in field. We define the angle between the basis vectors of the ZFC lattice and the basis vectors of an arbitrary lattice as the orientation, denoted $\phi$ in Fig.~\ref{Fig1}(d). Panels (b--d) show the VL at 0.2~T, but prepared after a decrease in magnetic field while held at 55~mK. The VL undergoes a clockwise rotation as the field is decreased, with a change in field of $-0.9$~T rotating the VL by around $25^{\circ}$. We refer to this as a rotation of the VL, although we must point out that SANS is unable to distinguish a local re-orientation of vortex nearest neighbors from a bulk rotation of the VL as a whole.

\begin{figure}
	\includegraphics[width=1\linewidth]{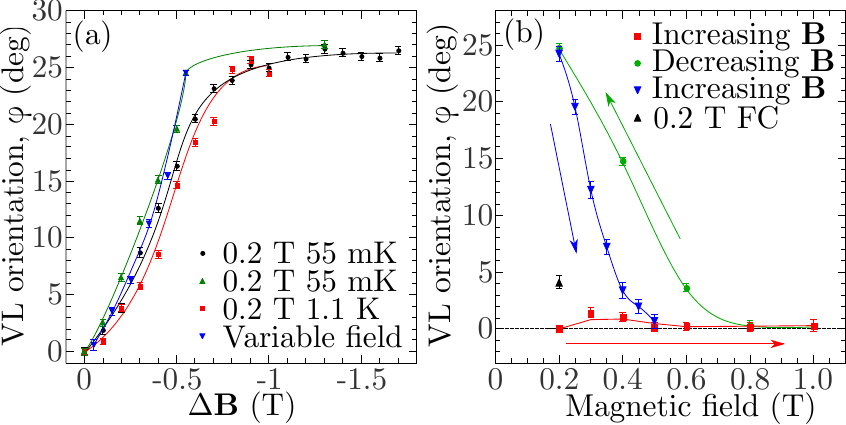}
	\caption{(a) Orientation of the VL, $\phi$, as a function of the change in magnetic field prior to measurement. (b) Orientation of the VL, $\phi$, as a function of absolute magnetic field at 55~mK. The orientation of the field-cooled 0.2~T lattice is also shown for reference.  Lines are guides for the eyes.}
	\label{Fig2}
\end{figure}

Figure~\ref{Fig2} presents a numerical representation of the rotation of the VL for magnetic field applied parallel to the \textbf{a} axis. Panel (a) shows the rotation of the VL as a function of the decrease in magnetic field prior to measurement. The initial magnetic field was applied from zero at 55~mK, raised to the relevant value and then decreased by the amount indicated on the graph. Two of the data sets, taken during separate experiments, were both measured at 0.2~T and 55~mK and correspond to the data in Fig.~\ref{Fig1}. They show the same behavior, although with a slight change in the rate of rotation which is probably due to a small difference in the alignment of the magnetic field with respect to the crystal axes, as all other experimental conditions were the same, and later data presented here indicate that the VL rotation is dependent on this alignment. These measurements were repeated at 1.1~K, which is $\sim 0.42~ T / T_{\rm c}$, and show the same behavior, indicating that the rotation is temperature independent. In the final measurement, labeled `variable field', the diffraction patterns were taken at different fields, but the starting field before the decrease prior to measurement was kept the same: 0.75~T.

Figure~\ref{Fig2}(b) plots the orientation of the VL as a function of magnetic field, with all measurements taken sequentially while remaining at 55~mK, starting at 0.2~T. The order of measurement is indicated by the arrows and follows the legend from the top down. As the magnetic field is increased from the 0.2~T ZFC lattice, the VL remains in the equilibrium state orientation up to the highest measured field of 1~T. Following this, the magnetic field is decreased and we see the lattice begin to rotate, reaching an angle of $25^{\circ}$ after returning to 0.2~T. The magnetic field was then increased again, and the lattice was seen to rotate back in an anticlockwise direction much faster than the initial clockwise rotation, returning to its initial orientation at around 0.5~T. Therefore, it appears that changing magnetic field below $T_{\rm c}$ always induces a rotation of the VL, however this rotation has saturation points at both 0 and 26 degrees. The orientation of the 0.2~T field-cooled lattice is shown, indicating that it has a slightly different orientation to the 0.2~T ZFC lattice. We hypothesize that this equilibrium state is very close to the 0 degree saturation point, and the act of raising field in the ZFC procedure is what causes the difference in orientations, since the previous observation indicates that this must be attempting to induce a counter-clockwise rotation.

\begin{figure}
	\includegraphics[width=1\linewidth]{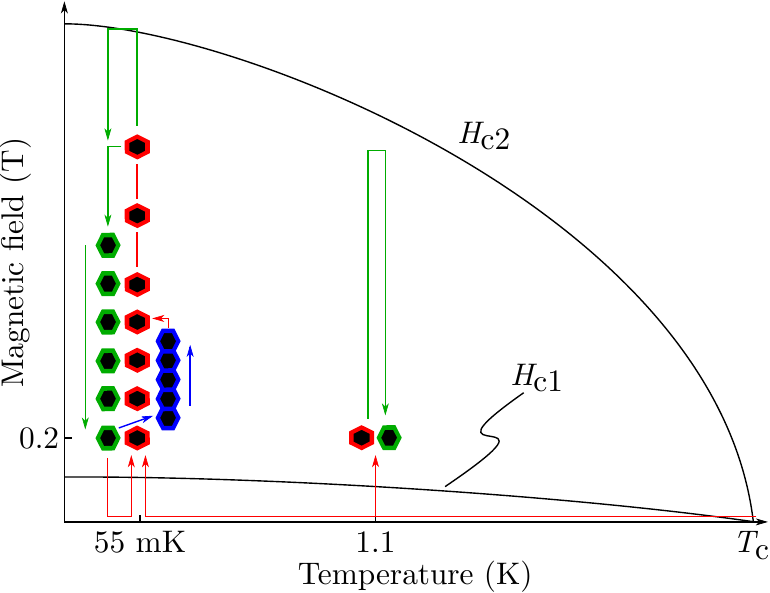}
	\caption{Schematic phase diagram indicating the data presented in this paper and the field/temperature paths used to obtain them, where the $H_{\rm c1}$ and $H_{\rm c2}$ lines are guides for the eye. Three separate sets of hexagons are shown, indicating the non-rotated (red), rotated by decreasing field (green) and rotated by increasing field (blue) VL orientations.}
	\label{Fig2a}
\end{figure}

Figure~\ref{Fig2a} presents a schematic phase diagram, showing the measurements presented in this paper for fields parallel to the \textbf{a} axis, and the paths taken in field and temperature used to prepare the VL. Three categories of VL are indicated: the equilibrium orientation, the lattice rotated by decreasing field and the lattice rotated by increasing field. 

\begin{figure}
	\includegraphics[width=1\linewidth]{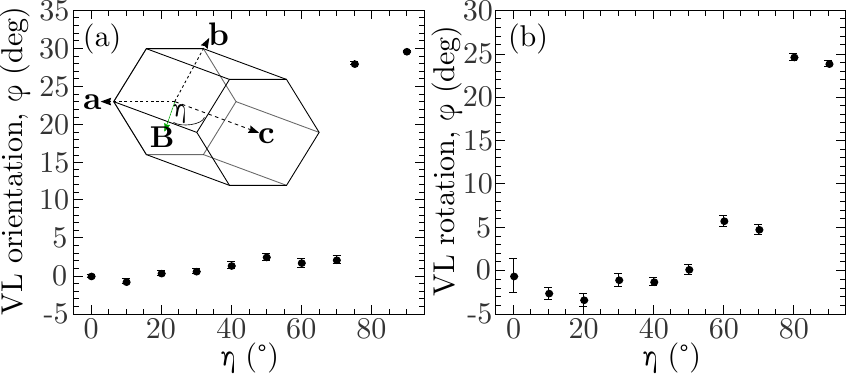}
	\caption{(a) The orientation of the VL as a function of the angle, $\eta$, between the applied magnetic field and the \textbf{c}-axis of the crystal. The VL orientation was measured with respect to the equilibrium state VL in the basal plane ($\mathbf{B} \parallel \mathbf{c}$). The inset illustrates a unit cell of the crystal and the orientation of the magnetic field. (b) Rotation of the VL, after a $-0.8$~T change in applied magnetic field, as a function of the angle between the magnetic field and the basal plane of the crystal. The orientation of the VL was measured with respect to the 0.2~T ZFC lattice at the same angle between the field and basal plane. All measurements were taken at 0.2~T and 55~mK.}
	\label{Fig3}
\end{figure}

Figure~\ref{Fig3} describes the orientation and rotation of the VL as a function of the angle $\eta$ between the applied magnetic field and the \textbf{c}-axis. Panel (a) plots the orientation of the ZFC VL at 0.2~T and 55~mK, measured with respect to the same conditions with the magnetic field applied parallel to the \textbf{c}-axis. The inset shows an illustration of the unit cell defining the angle $\eta$ between the magnetic field and the \textbf{c}-axis. Panel (b) shows the rotation of the VL after a $-0.8$~T change in magnetic field, measured with respect to the 0.2~T ZFC lattice at the same angle of magnetic field, $\eta$. We see in panel (a) that between $\eta = 70^{\circ}$ and $\eta = 75^{\circ}$ there is a reorientation of the VL of around $30^{\circ}$, and that rotation of the VL as a function of changing field below $T_{\rm c}$ emerges across this reorientation as the angle $\eta$ approaches $90^{\circ}$ where the field is in the \textbf{ab} plane.

\section{Discussion}

The observed rotation is clearly very unusual, and to illustrate this we will briefly compare it to the VL structural transitions in centrosymmetric systems. As discussed in the introduction, it is both predicted in theory and found in experiment that changes in the VL morphology have single-valued behavior as the field and temperature is varied~\cite{Kog97a, Kog97b, Fra97, Ich99, Nak02, Suz10}. Furthermore, at least in the absence of strong vortex pinning, it is a general result that the VL structure is independent of the thermodynamic path taken to produce it, a conclusion supported by an experiment, for example, on NbSn where the VL structure was found to be the same for both field cooled and ZFC preparations~\cite{Pau09}. This is in stark contrast to the behavior we observe in Ru$_{7}$B$_{3}$, where at no point can a typical structural transition be defined, but rather it is the process of changing magnetic field which acts as the driving force of the rotation. This is illustrated explicitly in Fig.~\ref{Fig2}(a) when comparing the `0.2~T 55 mK' scans with the `variable field 55 mK' data, which show the same rotation of the VL for equal changes in magnetic field whose paths in terms of absolute magnetic field have no overlap. We therefore conclude that the behavior we report here is due to other physics not yet explored by these models.

The reorientation of the VL as a function of $\eta$ in Fig~\ref{Fig3}(a), however, is an example of a conventional VL structure transition taking place. While the exact behavior of this transition is governed by local physics, as described by the theories discussed in the preceding paragraph, the existence of a transition as a function of $\eta$ is demanded by the geometric arguments of the `hairy ball theorem', which describes how continuous vector fields map onto the surface of objects with various topologies~\cite{Lav10}. In this case, the vectors describing the VL are mapped onto the sphere of possible orientations of the magnetic field with respect to the crystal, and because a sphere has an Euler characteristic of $\chi = +2$, this demands that there be singularities on the surface of the sphere which have a total winding number equal to $+2$. The symmetry of the VL allows for the existence of fractional singularities, and the existence of such a singularity along one of the six principal directions of the stereographic projection of the crystal suggests that we observe a $w = +1/6$ singularity~\cite{Lav10}. The dependence of the VL orientation on the field history of the sample adds an additional dimension to this problem, however, and it remains to be seen what effect this rotation will have on the location and nature of the singularities demanded by the hairy ball theorem.

One possible explanation for the rotation of the VL is the Magnus force, which in high $T_{\rm c}$ systems is responsible for phenomena such as the Hall effect and quantum vortex nucleation \cite{Bla94, Son97}. Considering the VL under the application of a changing field, it is evident that the trajectory of vortices must, on average, follow a radial path as the density of vortices throughout the sample is changed. Therefore, the direction of the Magnus force acting on a vortex is opposite to that on a corresponding vortex on the other side of the sample, leading to a net torque. To determine whether the Magnus force is a plausible explanation, we can estimate its strength and compare this to the pinning forces acting on the flux lines. We follow the expression from Ref.~\cite{Son13} for the Magnus force in a continuous superfluid at $T = 0$, $F_{\rm M} = m_{e} n v_{s} \times \mathbf{\zeta}$, where $n \approx 1 \times 10^{22}$ cm$^{-3}$ is the charge carrier density \cite{Fan09}, $m_{e}$ is the effective charge carrier mass, $v_{s}$ is the relative velocity between the flux lines and the superfluid, and the circulation quantum $\zeta = h / m_{e}$. The velocity of flux lines was estimated to be $1 \times 10^{-4}$~m$\cdot$s$^{-1}$ at the surface of a cylindrical superconductor of radius 2.5~mm, which is a reasonable approximation for the geometry of our sample, at a rate of change of magnetic field of $\sim 1$~T min$^{-1}$. This gives an order of magnitude estimate of the maximum Magnus force density within the sample to be $10^{5}$ N$\cdot$m$^{-3}$. We compare this to the pinning force density as estimated by the collective pinning model \cite{Lar79}, which allows us to determine pinning forces from the disorder in the VL as measured by SANS. From the diffraction data we extract the longitudinal correlation length, $\xi_{\rm L}$, and the radial correlation length, $\xi_{\rm R}$, which are calculated from the rocking curve width and radial spot width~\cite{Yar94, Pau12}. Throughout the field range where the rotation was observed, $0.2 - 0.8$~T, our diffraction data gave a rocking curve width of $\sim 0.2^{\circ}$ with a resolution of $\sim 0.15^{\circ}$, and a radial spot width $\omega_{\rm r}$ of $\sim 0.0007$ \AA$^{-1}$ with a resolution of $\approx 0.0005$ \AA$^{-1}$~\cite{Cub92, Pau12}. The correlation lengths from the SANS data are rescaled according to $L_{\rm c} , R_{\rm c} \approx \xi_{\rm L, R} (\xi / d_{\rm FLL})^{2}$ \cite{Gia95}, where $\xi$ is the coherence length, resulting in $L_{\rm c} \approx 10^{-6}$~m and $R_{\rm c} \approx 10^{-7}$~m. These are related to the pinning force density through the expressions 
$\rho_{\rm F} = 2 C_{44} \xi / L_{\rm c}^2$ and $\rho_{\rm F} = C_{66} \xi / 4 R_{\rm c}^{2}$,
where we follow the London and thermodynamic limits for the shear modulus $C_{66} = \bar{B}B_{\rm c2} / (8 \kappa ^2 \mu_{0} )$ and tilt modulus $C_{44} = B^2 / \mu_{0}$ \cite{Bran10}, $\kappa$ being the Ginzburg-Landau parameter. Both methods gave a pinning force density of $\rho_{\rm F} \approx 10^{9}$~N$\cdot$m$^{-3}$. This is four orders of magnitude higher than our estimate for the maximum Magnus force. However, we note that the Magnus force itself is not the depinning force, as the vortices must already be depinned by the change in magnetic field, although this calculation suggests that it would not play a significant role in the vortex dynamics. Furthermore, we expect the Magnus force to be independent of field orientation, and the rocking curve width showed no change with $\eta$ outside of error, indicating a correspondingly direction-independent pinning, whereas the rotating behavior only appears when the field is close to the \textbf{a}-axis. Finally, the Magnus force is linearly dependent on the carrier density, which varies with temperature in a superconductor, whereas no temperature dependence is observed in Fig.~\ref{Fig2}(a). This leads us to conclude that the Magnus force is not responsible for the rotation of the VL.

Recently Ru$_{7}$B$_{3}$ was studied using  $\mu$SR~\cite{Par14}, and these measurements reveal the presence of spontaneous magnetic fields below the superconducting transition temperature which indicate that the superconducting state breaks time-reversal symmetry. In turn this gives rise to the problem of whether the time-reversal symmetry breaking (TRSB) superconducting state together with the broken inversion symmetry can drive the rotation of the VL. To address this possibility, we consider an extended Ginzburg-Landau (GL) approach by adding a magnetic contribution from the spontaneous magnetization 
\begin{equation}
F = F_s + F_m - \frac{{{{\bf{B}}^2}}}{{8\pi }} - {\bf{B}} \cdot {\bf{M}}.
\label{M+GL}
\end{equation}
Here, ${F_s}$ is the superconducting part of the energy, which takes into account the uniaxial symmetry of the crystal and the absence of inversion symmetry for the ${C_{6v}}$ point group in the presence of spin-orbit coupling. With $D = \left( { - i\hbar \nabla  - \frac{{2e}}{{\hbar c}}{\bf{A}}} \right)$, we write
\begin{eqnarray}
F_s = \int {\left\{ {\alpha {{\left| \psi  \right|}^2} + \frac{1}{2}\beta {{\left| \psi  \right|}^4} + \gamma {{\left| D\psi \right|}^2}} \right.} \nonumber \\
+ ~ \left. {{\rm{           }}\varepsilon {\bf{n}} \cdot {\bf{B}} \times \left[ {{\psi ^ * }D\psi  + \psi D^*{\psi ^ * }} \right]} \right\}dV,
\label{GL_superconductor2}
\end{eqnarray}
where $\alpha$ and $\beta$ are phenomenological constants, $\gamma$ is related to the average Fermi velocity, the term containing $\varepsilon {\bf{n}} \cdot {\bf{B}} \times \textbf{j}$ is the so-called Lifshitz invariant arising from the loss of inversion symmetry, $\bf{n}$ is a vector parallel to the sixfold rotation axis \textbf{c}, and $\varepsilon$ is a parameter related to the strength of the spin-orbit coupling. We represent the possibility of spontaneous magnetic fields from TRSB by including a magnetic part of the free energy $F_m$, in a similar manner to a previous study focusing on the emergence of spontaneous magnetic fields at twin boundaries in NCS superconductors~\cite{Ach14}, which we define as
\begin{equation}
F_m = \int {\left\{ {a{{\left| m \right|}^2} + \frac{1}{2}b{{\left| m \right|}^4} + {d_{ij}}{\nabla _i}m{\nabla _j}m} \right\}}dV,
\label{magnetic_energy}
\end{equation} 
where $m$ is the density of the magnetic moment component. 

Following Abrikosov's procedure~\cite{Abr57}, which is written out in detail in the appendix, we find an expression for the magnetization $M$ of a superconductor below the upper critical field ${H_{c2}}$
\begin{widetext}
\begin{eqnarray}
\frac{{M - {m_0}}}{{H - {H_{c2}}}} = \frac{{\left( {a + 3bm_0^2} \right){{\left( {\frac{{2\pi }}{{{\Phi _0}}}\gamma  + 2\varepsilon } \right)}^2}}}{{2\beta \left( {a + 3bm_0^2 - 2\pi } \right) + 4\pi {{\left( {\frac{{2\pi }}{{{\Phi _0}}}\gamma  + 2\varepsilon } \right)}^2}\left( {a + 3bm_0^2} \right)}}\frac{{{{\left\langle {{{\left| {{\psi _0}} \right|}^2}} \right\rangle }^2}}}{{\left\langle {{{\left| {{\psi _0}} \right|}^4}} \right\rangle }},
\label {magnetization}
\end{eqnarray}
\end{widetext}
where $m_0$ is the density of the magnetic moment of the spontaneous magnetization and the brackets $\left\langle {...} \right\rangle$ define the spatial average of the order parameter. We should note that the right side in Eq.~(\ref{magnetization}) is inversely proportional to the Abrikosov parameter, ${\beta _A}$. The minimal value of ${\beta _A}$  corresponds to a global minimum of the GL free energy for a superconductor and determines the energetically favorable VL configuration. For a conventional single-band superconductor the Abrikosov parameter is material independent and predicts a hexagonal lattice with a degenerate orientation with respect to the crystal. In our case, however, we can see from Eq.~(\ref{magnetization}) that for a non-centrosymmetric superconductor with TRSB this universality of $\beta_A$ is lost, and the configuration of the lattice starts to depend on the the external magnetic field, microscopic magnetization, and the superconducting properties of the system:
\begin{widetext}
\begin{equation}
\beta _A \sim \frac{{2\beta \left( {a + 3bm_0^2 - 2\pi } \right) + 4\pi {{\left( {\frac{{2\pi }}{{{\Phi _0}}}\gamma  + 2\varepsilon } \right)}^2}\left( {a + 3bm_0^2} \right)}}{{\left( {a + 3bm_0^2} \right){{\left( {\frac{{2\pi }}{{{\Phi _0}}}\gamma  + 2\varepsilon } \right)}^2}}}\frac{{\left\langle {{{\left| {{\psi _0}} \right|}^4}} \right\rangle }}{{{{\left\langle {{{\left| {{\psi _0}} \right|}^2}} \right\rangle }^2}}}.
\label{AP}
\end{equation}
\end{widetext}

\begin{figure}
\includegraphics[width=1\columnwidth]{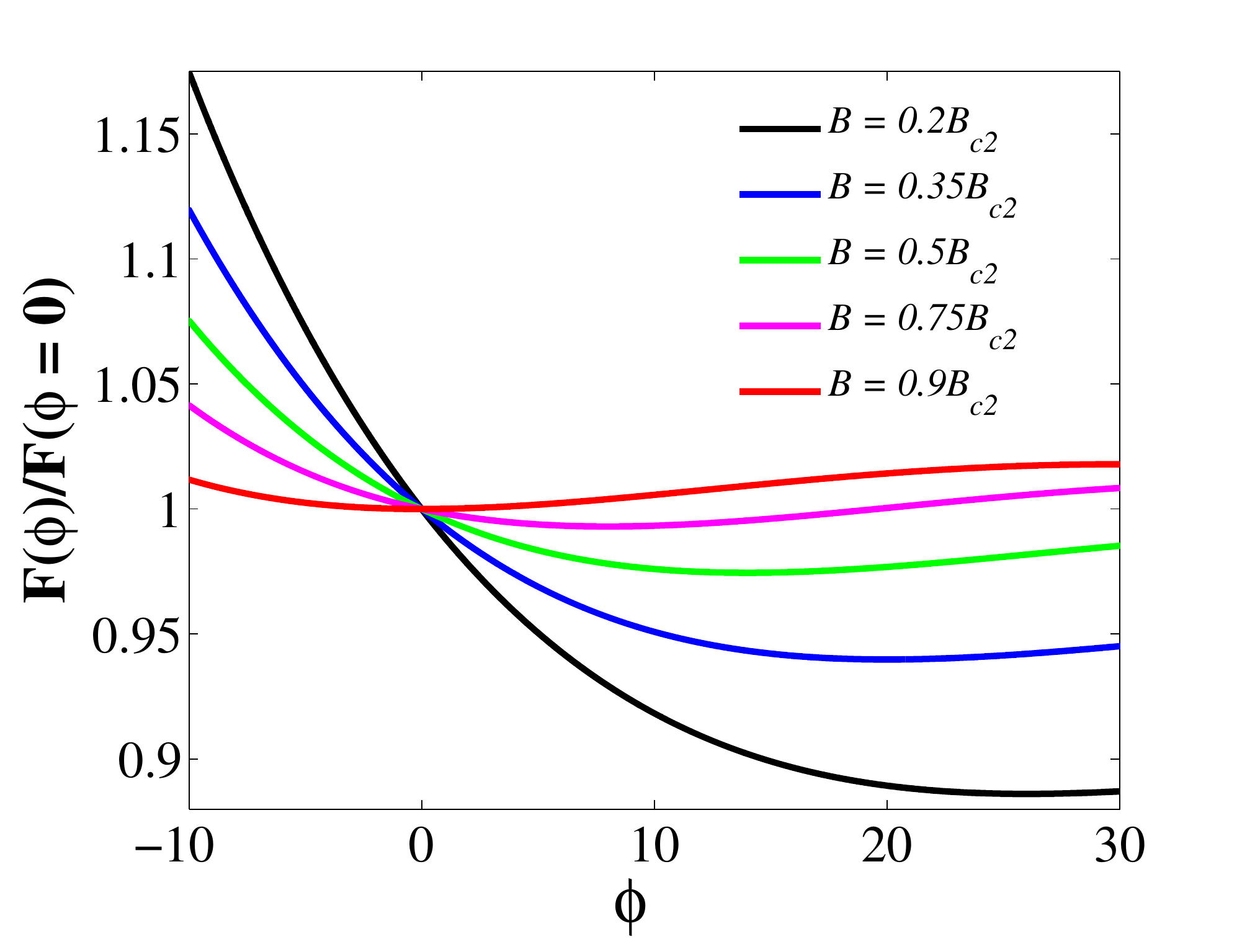}
\caption{Free energy of VL vs. the rotation angle as a function of magnetic field, for the GL model described in the text.}
\label{figS1}
\end{figure}

Therefore, from the determination of the value and behavior of the spontaneous magnetization  ${m_0}$ and knowing the exact values of phenomenological parameters (which can be extracted from the microscopic description) for the given superconducting compound we can calculate $\beta _A$ and then the corresponding free energy. To illustrate the coupling of the VL to the orientation of the crystal lattice we performed calculations of the free energy based on Eq.~(\ref{GL_energy_final}) as a function of the angle $\phi$ between the basis vectors of the VL and the crystal axes assuming an arbitrary value of $m_{0}$, where the orientation of the VL $\phi$ appears from the averaging procedure for the order parameter within the unit cell of the lattice~\cite{Jam69, Hir13}. An angle of $\phi = 0$ corresponds to the equilibrium state orientation of the VL which we see in increasing field. Using a trial set of numerical parameters: $a=-2.3$, $b= 0.75$, $\beta=2$, $\gamma=3.7$ and $\varepsilon=1.4$, we plot Fig.~\ref{figS1} which shows that the coupling of the VL to the broken inversion symmetry and spontaneous magnetization of TRSB gives the VL a field-dependent orientation. While these parameters are related to microscopic properties of the system, not all of them are currently known, and so the above were chosen purely to illustrate the rotation of the VL which we observed. In order to correctly model the hysteretic orientation of the VL it will be necessary to know explicit values of these parameters, the spontaneous magnetization, and how these interact with the more complex vortex structures of NCS superconductors.

Our model demonstrates that both the spontaneous fields present under TRSB and the effect of broken inversion symmetry inseparably couple to the orientation and structure of the VL. Since the rotation of the VL is only observed when the field is close to the \textbf{ab}-plane, and the Lifshitz invariant, which is proportional to ${\bf{n}} \cdot {\bf{B}} \times \textbf{j}$, drops out of the free energy when the applied magnetic field is parallel to the \textbf{c} axis, the model suggests that the effect of broken inversion symmetry is necessary for the rotation to appear. The model also predicts that the spontaneous moments which arise from TRSB are also coupled to the VL orientation. These should be randomly oriented in zero field, and the hysteresis in the VL orientation suggests that they are aligned by the applied magnetic field and undergo a hysteresis of their own, driving the changes is the VL orientation. The alignment and hysteresis of TRSB fields by an external applied field has already been observed in the heavy-fermion superconductor UPt$_{3}$~\cite{Sch14}, and while the TRSB fields are very small, on the order of 1~Gauss, we note first that the the orientation and coordination of the VL is notoriously sensitive to small changes in its free energy~\cite{Kle64}, and second that superconducting states which break time reversal symmetry tensorially couple the supercurrent to gradients in the order parameter, resulting in additional field components within the mixed state~\cite{Sig91}. This is similar to the appearance of tangential fields within the vortex lattice of NCS superconductors, although its effects have not been studied and it may mean that the TRSB fields are not limited to those observed in zero field by techniques such as $\mu$SR. Furthermore, it has also been found that superconducting order parameters which break time-reversal symmetry can have a significant effect on the VL through other mechanisms, forming vortex cores which break rotational symmetry and result in frustrated lattices~\cite{Tok90}, and so we therefore consider the effect of TRSB to be relevant despite the small size of the spontaneous field it produces.

The GL approach we use here should be considered a qualitative model, to illustrate the coupling of NCS superconductivity and TRSB to the behavior of the VL. At present, the model demonstrates not only the presence of the rotation but also captures its anisotropic nature; that is to say that it only occurs in one direction. Furthermore, the presence of a shifting minimum in the free energy as a function of $\phi$ illustrates why the rotated states appear so stable to both perturbations in field and thermal fluctuations, as they are new equilibrium states as opposed to metastable states. However, before the model can truly capture this behavior, it requires several improvements. Most notably, other contributions to $F(\phi)$, such as Fermi velocity anisotropy and gap anisotropy, must be included, as while the Abrikosov parameter predicts a degenerate orientation for a conventional superconductor~\cite{Abr57}, this is never observed and the VL always chooses a specific orientation due to these physics which are not captured by the original model. It is not yet known what effect the broken inversion symmetry will have on the results of theories developed after Abrikosov's, which have been used to explain the VL orientation and coordination in centrosymmetric systems, which were discussed earlier, if any. Our phenomenological approach also did not take into account the possible presence of a small triplet component of the order parameter, as Ru$_{7}$B$_{3}$ has been suggested to be a pure \textit{s}-wave system from magnetization measurements~\cite{Fan09}, although the data were not taken down to a low enough fraction of $T / T_{\rm c}$ to be certain. In this case the expression for the Abrikosov parameter will be more complex, allowing for substantially richer behavior of the VL. In order to develop a more complete understanding of the VL behavior in this system, it will be necessary to perform microscopic calculations which include details of the anisotropy in the Fermi velocity and the superconducting gap as well as the ASOC and TRSB. Furthermore, it is imperative to investigate other NCS and TRSB superconductors under these conditions to further elucidate the contributions of TRSB and broken inversion symmetry to the VL behavior we observe here. 

\section{Conclusions}

We have performed SANS measurements on the VL of the noncentrosymmetic superconductor Ru$_{7}$B$_{3}$, finding a VL orientation which is strongly dependent on the field history of the sample within the superconducting state. This is unprecedented behavior, which has not been predicted by previous theories of the VL. To address this, we construct a model of the VL from the phenomenological GL theory which includes the Lifshitz invariant suitable for our material and the magnetic contribution of the spontaneous magnetization due to TRSB phenomena. We find that the Abrikosov parameter, a geometric object which relates to the orientation and coordination of the VL, gains a complex pre-factor in the case of a noncentrosymmetic superconductor with TRSB, which couples the parameter to the spontaneous magnetization and superconducting properties of the system. We therefore predict that the spontaneous magnetization has hysteretic behavior in Ru$_{7}$B$_{3}$, which in turn results in a corresponding hysteresis in the VL orientation.

\section{Acknowledgements}

This project was funded by the German Research Foundation (DFG) through the research grants IN 209/3-2, IN 209/6-1, and the Graduiertenkolleg GRK 1621. A.S.S. acknowledges support from the International Max Planck Research School for Chemistry and Physics of Quantum Materials (IMPRS-CPQM). The work at Warwick was supported by EPSRC, UK, through Grant EP/M028771/1. 

\section{Appendix}

Our approach is based on a Ginzburg-Landau (GL) functional 
\begin{equation}
F = {F_s} + {F_m} - \frac{{{{\bf{B}}^2}}}{{8\pi }} - {\bf{B}} \cdot {\bf{M}},
\label{GL_total}
\end{equation} 
with the terms as defined previously and the Lifshitz invariant which describes the ASOC given in Eq.~\ref{GL_superconductor2}. It's important to note that the contribution to the GL energy of the Lifshitz invariant gives rise to different phenomena in noncentrosymmetric superconductors, in particular FFLO-like phases, magnetoelectric effects and exotic vortex states.

We also introduce a magnetic part of the free energy from the spontaneous magnetization due to TRSB, which is defined as 
\begin{equation}
{F_m} = \int {\left\{ {a{{\left| m \right|}^2} + \frac{1}{2}b{{\left| m \right|}^4} + {d_{ij}}{\nabla _i}m{\nabla _j}m} \right\}},
\label{magnetic_part}
\end{equation}
where $m$ is the density of the magnetic moment component. 
By minimizing the GL free energy with respect to the order parameter and the vector potential the following GL equations are found
\begin{gather}
\alpha \left| \psi  \right| + \beta {\left| \psi  \right|^3} + \gamma {\left( { - i\hbar \nabla  - \frac{{2e}}{{\hbar c}}{\bf{A}}} \right)^2}\psi  \nonumber \\
+ ~ \varepsilon {\bf{n}} \cdot {\bf{B}} \times \left( { - i\hbar \nabla  - \frac{{2e}}{{\hbar c}}{\bf{A}}} \right)\psi  = 0,
\label{GL_OP}
\end{gather}
and
\begin{gather}
{\bf{j}} = \frac{{2ie}}{{\hbar c}}\gamma \left[ {{\psi ^ * }\left( { - i\hbar \nabla  - \frac{{2e}}{{\hbar c}}{\bf{A}}} \right)\psi  + \psi \left( {i\hbar \nabla  - \frac{{2e}}{{\hbar c}}{\bf{A}}} \right){\psi ^ * }} \right] \nonumber \\ 
+ ~\frac{{4e}}{{\hbar c}}\varepsilon {\left| \psi  \right|^2}{\bf{B}}.
\label{GL_current}
\end{gather}
For simplicity we choose the gauge ${\bf{A}} = Bx\left( {0,1,0} \right)$.  Near the upper critical field we can linearize the GL Eq.~(\ref{GL_OP})
\begin{eqnarray}
\alpha \psi  - \gamma \left( {\frac{{{\partial ^2}}}{{\partial {x^2}}} + {{\left( {\frac{{{\partial ^2}}}{{\partial {y^2}}} - \frac{{2ie{B_{c2}}x}}{{\hbar c}}} \right)}^2}} \right)\psi  \nonumber \\ +
~\varepsilon {B_{c2}}\left( {\frac{\partial }{{\partial x}} + \left( {\frac{\partial }{{\partial y}} - \frac{{2ie{B_{c2}}x}}{{\hbar c}}} \right)} \right)\psi  = 0.
\label{GL_OP_lin}
\end{eqnarray}
The lowest eigenvalue of the GL operator corresponds to the order parameter
\begin{eqnarray}
\psi \left( {x,y} \right) = \sum\limits_{n =  - \infty }^\infty  {{C_n}\exp \left( {ikny} \right)} \nonumber \\ 
\times \exp \left[ { - \frac{{\pi {B_{c2}}}}{{{\Phi _0}}}{{\left( {x - \frac{{kn{\Phi _0}}}{{2\pi {B_{c2}}}} - \frac{{{\Phi _0}}}{{2\pi }}\frac{\varepsilon }{\gamma }} \right)}^2}} \right].
\label{Eugenvalue}
\end{eqnarray}
Eq.~(\ref{Eugenvalue}) for the vortex lattice solution coincides with that of Abrikosov for a single-band superconductor, but due to the presence of the Lifshitz invariant it obtains the so-called helical phase factor 
$\frac{{{\Phi _0}}}{{2\pi }}\frac{\varepsilon }{\gamma }$, which can lead to the enhancement of the upper critical field.
At a magnetic field $B$ slightly below $B_{c2}$ the mixed state appears and the order parameter amplitude, the magnetic moment, and the vector-potential acquire the small correction
\begin{equation}
\psi  = {\psi _0} + {\psi _1}, {\bf{m}} = {{\bf{m}}_{\bf{0}}} + {{\bf{m}}_{\bf{1}}}, {\bf{A}} = {{\bf{A}}_{\bf{0}}} + {{\bf{A}}_{\bf{1}}},
\label{corrections}
\end{equation}
where ${{\bf{A}}_{\bf{1}}} = \left( {0,\left( {H - 4\pi {m_0} - {B_{c2}}} \right)x,0} \right) + \delta {\bf{A}}$, $\mathbf{m}_0$ is the spontaneous magnetization, ${{\bf{A}}_{\bf{0}}} = Bx\left( {0,1,0} \right)$, and $\mathbf{m}_{1}$ and $\psi_{1}$ are the corrections due to the presence of the mixed state to the magnetization and order parameter respectively.
The corresponding magnetic induction is 
\begin{equation}
{\bf{B}} = {\bf{H}} + 4\pi {{\bf{m}}_{\bf{0}}} + \delta {\bf{B}}.
\label{mag_ind}
\end{equation}
The full magnetization of the system is 
\begin{equation}
{\bf{M}} = \frac{{\left\langle {\left( {{\bf{B}} - {\bf{H}}} \right)} \right\rangle }}{{4\pi }} = {{\bf{m}}_{\bf{0}}} + \frac{{\left\langle {\delta {\bf{B}}} \right\rangle }}{{4\pi }},
\label{full_mag}
\end{equation}
where brackets $\left\langle {...} \right\rangle $ define the spatial average.
Taking into account Eq.\ (\ref{GL_current}) one can get an expression for the current in the form
\begin{gather}
{\bf{j}} = \frac{1}{{4\pi }}{\bf{rot}}\left( {\delta {\bf{B}} - 4\pi {{\bf{m}}_{\bf{1}}} - 8\pi \varepsilon {\bf{n}}{{\left| {{\psi _0}} \right|}^2}} \right) = \nonumber \\ 
\gamma \left( {\psi _0^ * \left( { - i\hbar \nabla  - \frac{{2e}}{{\hbar c}}{{\bf{A}}_{\bf{0}}}} \right){\psi _0} + {\psi _0}\left( {i\hbar \nabla  - \frac{{2e}}{{\hbar c}}{{\bf{A}}_{\bf{0}}}} \right)\psi _0^ * } \right).
\label{new_current}
\end{gather}
Representing Eq.~(\ref{new_current}) through the $x$ and $y$ scalar components and taking into account the relation $\frac{{\partial \psi_0 }}{{\partial x}} = \left( { - i\frac{\partial }{{\partial y}} - \frac{{2ie{B_{c2}}}}{{\hbar c}}} \right)\psi _0$ one can rewrite 
\begin{equation}
\frac{\partial }{{\partial x}}\left( {\delta B - 4\pi {m_1} - 8\pi \varepsilon {{\left| {{\psi _0}} \right|}^2}} \right) = 4\pi \frac{{2\pi }}{{{\Phi _0}}}\gamma \frac{{\partial {{\left| {{\psi _0}} \right|}^2}}}{{\partial x}},
\label{new_current_x}
\end{equation}
and
\begin{equation}
\frac{\partial }{{\partial y}}\left( {\delta B - 4\pi {m_1} - 8\pi \varepsilon {{\left| {{\psi _0}} \right|}^2}} \right) = 4\pi \frac{{2\pi }}{{{\Phi _0}}}\gamma \frac{{\partial {{\left| {{\psi _0}} \right|}^2}}}{{\partial y}}.
\label{new_current_y}
\end{equation}
This gives the expression for the correction $\delta B$
\begin{equation}
\delta B = 4\pi \frac{{2\pi }}{{{\Phi _0}}}\gamma {\left| {{\psi _0}} \right|^2} + 4\pi {m_1} + 8\pi \varepsilon {\left| {{\psi _0}} \right|^2}.
\label{delta_B}
\end{equation}
To find ${{{m}}_{{1}}}$  we consider the variation of the free energy given by Eq.~(\ref{magnetic_part}) in respect to ${{m}}$
\begin{equation}
2am + 2b{m^3} + 2{d_{ij}}{\nabla ^2}m - B = 0,
\label{Eq_m}
\end{equation}
hence the correction ${{{m}}_{{1}}}$ for the magnetization is determined by the equation
\begin{equation}
\left( {2a + 6bm_0^2 + 2{d_{ij}}{\nabla ^2}} \right){m_1} - \delta B = 0.
\label{Eq_m_1}
\end{equation}
If we assume that the magnetic coherence length is smaller than the size of the vortex core then one can neglect the Laplacian term in the Eq.~(\ref{Eq_m_1}), and taking into account Eq.~(\ref{delta_B}) we get the expression for 
\begin{equation}
{m_1} = \frac{{\frac{{{{\left( {2\pi } \right)}^2}}}{{{\Phi _0}}}\gamma {{\left| {{\psi _0}} \right|}^2} + 4\pi \varepsilon {{\left| {{\psi _0}} \right|}^2}}}{{a + 3bm_0^2 - 2\pi }}.
\label{Eq_m_1_new}
\end{equation}
According to Eq.~(\ref{full_mag}) below transition to the superconducting state the magnetization follows 
\begin{equation}
{{M}} - {{{m}}_{{0}}} = \left\langle {\frac{{2\pi }}{{{\Phi _0}}}\gamma {{\left| {{\psi _0}} \right|}^2} + {m_1} + 2\varepsilon {{\left| {{\psi _0}} \right|}^2}} \right\rangle.
\label{full_mag_new}
\end{equation}
Now we proceed to find an average of Eq.~(\ref{full_mag_new}). Based on the linearized form of the GL Eq.~(\ref{GL_OP_lin}), after long but straightforward calculations we obtain 
\begin{equation}
\left\langle {{\bf{j}} \cdot {{\bf{A}}_{\bf{1}}} + 2\beta {{\left| {{\psi _0}} \right|}^4}} \right\rangle  = 0.
\label{average1}
\end{equation}
Using Eqs.~(\ref{new_current_x}) and~(\ref{delta_B}) for the current and small correction of $\delta B$ respectively we have  
\begin{widetext}
\begin{equation}
\left\langle {\left( {\frac{{2\pi }}{{{\Phi _0}}}\gamma {{\left| {{\psi _0}} \right|}^2} - 2\varepsilon {{\left| {{\psi _0}} \right|}^2}} \right)\left( {H - {H_{c2}}} \right) + \left( {\frac{{2\pi }}{{{\Phi _0}}}\gamma {{\left| {{\psi _0}} \right|}^2} - 2\varepsilon {{\left| {{\psi _0}} \right|}^2}} \right)\delta B + 2\beta {{\left| {{\psi _0}} \right|}^4}} \right\rangle  = 0.
\label{Abrikosov_identity}
\end{equation}

Thus below the upper critical field the magnetization decrease is
\begin{equation}
\left\langle {\left( {\frac{{2\pi }}{{{\Phi _0}}}\gamma {{\left| {{\psi _0}} \right|}^2} - 2\varepsilon {{\left| {{\psi _0}} \right|}^2}} \right)\left( {H - {H_{c2}}} \right) +\left( {\frac{{2\pi }}{{{\Phi _0}}}\gamma {{\left| {{\psi _0}} \right|}^2} - 2\varepsilon {{\left| {{\psi _0}} \right|}^2}} \right)\delta B + 2\beta {{\left| {{\psi _0}} \right|}^4}} \right\rangle  = 0,
\label{full_mag_2}
\end{equation}
and final expression for Eq.~(\ref{full_mag_2}) takes the form
\begin{equation}
M - {m_0} = \frac{{\left( {a + 3bm_0^2} \right){{\left( {\frac{{2\pi }}{{{\Phi _0}}}\gamma  + 2\varepsilon } \right)}^2}}}{{2\beta \left( {a + 3bm_0^2 - 2\pi } \right) + 4\pi {{\left( {\frac{{2\pi }}{{{\Phi _0}}}\gamma  + 2\varepsilon } \right)}^2}\left( {a + 3bm_0^2} \right)}}\frac{{{{\left\langle {{{\left| {{\psi _0}} \right|}^2}} \right\rangle }^2}}}{{\left\langle {{{\left| {{\psi _0}} \right|}^4}} \right\rangle }}\left( {H - {H_{c2}}} \right).
\label{full_mag_final}
\end{equation}
\end{widetext}
The pre-factor before the expression $(H - {H_{c2}})$  can be interpreted as the relation, known for a single-band superconductor, $1 / {{4\pi {\beta _A}\left( {2{k^2} - 1} \right)}}$, where $k$ is the GL parameter and ${\beta _A}$ is the Abrikosov parameter.

In a conventional, centrosymmetric single-band superconductor the averaging procedure for Eq.~(\ref{full_mag_final}) can be carried out analytically yielding the expression
\begin{equation}
{\beta _A} = \frac{{\left\langle {{{\left| {{\psi _0}} \right|}^4}} \right\rangle }}{{{{\left\langle {{{\left| {{\psi _0}} \right|}^2}} \right\rangle }^2}}} = \sqrt {\sin \theta } \left[ {{{\left| {\vartheta _3^2\left( {0,{e^{i\theta }}} \right)} \right|}^2} + {{\left| {\vartheta _2^2\left( {0,{e^{i\theta }}} \right)} \right|}^2}} \right],
\label{b_A_single_band}
\end{equation}
where ${\vartheta _2}\left( {z,\tau } \right)$ and ${\vartheta _3}\left( {z,\tau } \right)$ are Jacobi theta functions, and $\theta$ is the angle between the two basis vectors which define the vortex lattice. The value of an angle $\theta $ which minimizes ${\beta _A}$ defines the structure of the vortex lattice in a superconductor. In turn the minimal value of ${\beta _A}$ corresponds to the global minimum of a GL functional and for a conventional single-band superconductor it is reached for $\theta  = \pi /3$ (or  $\theta  = 2\pi /3$) regardless of the superconducting material. 

In the case of a noncentrosymmetric superconductor with TRSB this universality is lost through the introduction of the pre-factor in Eq.~(\ref{full_mag_final}), which couples the VL orientation to the crystal lattice through the TRSB moments and the ASOC. Based on Eq.~(\ref{Eugenvalue}) for the order parameter we can obtain the value of $\beta_{A}$ including modifications from Eq.~(\ref{full_mag_final}). Taking this into account, the free energy of a superconductor given by Eq.~(\ref{GL_total}) can be reduced to the form	
\begin{equation}
F = {F_m} + \frac{{{B^2}}}{{8\pi }} - \frac{{{{\left( {B - {B_{c2}}} \right)}^2}}}{{1 + {\beta _A}\left( {2{k^2} - 1} \right)}}.
\label{GL_energy_final}
\end{equation}

\bibliographystyle{revtest6}
\bibliography{Ru7B3}

\end{document}